\documentclass{emulateapj}

\def\ltsima{$\; \buildrel < \over \sim \;$}
\def\simlt{\lower.5ex\hbox{\ltsima}}
\def\gtsima{$\; \buildrel > \over \sim \;$}
\def\simgt{\lower.5ex\hbox{\gtsima}}

\newcommand{\solarmass}{\mbox{${\rm M_{\odot}}$}}

\newcommand{\nhone}{$N_{\rm H}^{\rm 1}$}
\newcommand{\nhtwo}{$N_{\rm H}^{\rm 2}$}
\newcommand{\nhg}{$N_{\rm H}^{\rm gal}$}
\newcommand{\cosmo}{($H_0$, $\Omega_{\rm m}$, $\Omega_{\lambda}$)}

\shorttitle{{\it Suzaku} Observation of 4C 50.55}
\shortauthors{Tazaki et al.}

\begin{document}

\title{{\it Suzaku} Observation of the Brightest Broad-Line Radio Galaxy 
4C~50.55\ (IGR J 21247+5058)}
\author{
 Fumie Tazaki\altaffilmark{1},
 Yoshihiro Ueda\altaffilmark{1},
 Yukiko Ishino\altaffilmark{1},
 Satoshi Eguchi\altaffilmark{1},
 Naoki Isobe\altaffilmark{1},
\\
 Yuichi Terashima\altaffilmark{2},
 Richard F. Mushotzky\altaffilmark{3}
}

\altaffiltext{1}{Department of Astronomy, Kyoto University, Kyoto 606-8502, Japan}
\altaffiltext{2}{Department of Physics, Ehime University, Matsuyama 790-8577, Japan}

\altaffiltext{3}{Department of Astronomy, University of Maryland, College Park, MD, USA}

\begin{abstract} 

We report the results from a deep {\it Suzaku} observation of 4C 50.55
(IGR J 21247+5058), the brightest broad-line radio galaxy in the hard
X-ray ($>$ 10 keV) sky. The simultaneous broad band spectra over 1--60
keV can be represented by a cut-off power law with two layers of
absorption and a significant reflection component from cold matter
with a solid angle of $\Omega/2\pi \approx 0.2$. A rapid flux rise by
$\sim 20$\% over $2\times10^4$ sec is detected in the 2--10 keV
band. The spectral energy distribution suggests that there is little
contribution to the total X-ray emission from jets. Applying a thermal
Comptonization model, we find that corona is optically thick
($\tau_{\rm e} \approx 3$) and has a relatively low temperature
($kT_{\rm e} \approx 30 $ keV). The narrow iron-K emission line is
consistent with a picture where the standard disk is truncated and/or
its inner part is covered by optically thick Comptonizing corona
smearing out relativistic broad line features. The inferred disk
structure may be a common feature of accretion flows onto black holes
that produce powerful jets.

\end{abstract}


\section{Introduction}

Radio galaxies are a class of Active Galactic Nuclei (AGNs) spouting
powerful radio jets whose axis is not aligned along the line of
sight. While the formation mechanism of relativistic jets is not fully
understood, it must be closely related to mass accretion flow onto the
central supermassive black hole. In fact, studies of Galactic black
holes have revealed that the relative power of the jets to accretion
critically depends on the state of the accretion disk, which is
predominantly determined by the mass accretion rate \citep[e.g., see]
[]{Fen04b}. For AGNs, however, detailed studies of accretion disk state
in relation to the jet formation are still limited. Broad band
observation of nearby, bright radio galaxies hence give us ideal
opportunities to unveil this problem, because, unlike ``blazars'', the
innermost disk can be well observable in the X-ray band with much
smaller contribution from the jet emission.

4C 50.55 (IGR J21247+5058) is a Fanaroff-Riley type II broad line
radio galaxy (BLRG), discovered by {\it INTEGRAL} in its first survey
catalogue \citep{Bir04}. The source is also detected in the first 9
months data of {\it Swift} Burst Alert Telescope (BAT) survey of AGNs
\citep{Tue08}. Though being the brightest BLRG in the hard X-ray sky
above 10 keV, 4C 50.55, located at $(l,b) = (93.32,0.3937)$, had been
unrecognized as a bright X-ray source until the {\it INTEGRAL} and
{\it Swift} era, due to the obscuration by the Galactic plane. This
source is also of great interest for understanding the accretion flow
onto supermassive black holes at high fractions of Eddington
luminosity, which is estimated to be $L_{\rm bol}/L_{\rm Edd} \sim
0.4$ (see section~\ref{differ_SED}).  \citet{Mas04} determined the redshift
to be $z = 0.020 \pm0.001$ based on the detection of a broad
H${\alpha}$ emission line in the optical spectrum. From the observed
flux densities at 1.4 GHz with the Very Large Array (VLA),
\citet{Mol07} estimate the inclination angle to be $\theta \sim
35^{\circ}$, assuming a Lorentz factor of $\gamma = 5$ and a typical
power ratio between the core and total known for radio galaxies
\citep{Gio88}. The inferred viewing geometry of the nucleus is well
consistent with the classification of 4C 50.55 as a BLRG.

The initial results on the X-ray spectra are reported by
\citet{Mol07}, who used the {\it XMM-Newton} and {\it Swift}/XRT data
combined with a time-averaged {\it INTEGRAL} spectrum. They find that
the spectra below 10 keV are apparently hard, and complex, multiple
layers of absorber are required. Iron-K features are not significantly
detected and the strength of reflection components, $R \equiv
\Omega/2\pi$, where $\Omega$ is the solid angle of the reflector, is
not tightly constrained ($R \simlt 0.9$). In this paper, we present
the first simultaneous broad band X-ray data of 4C 50.55 observed with
{\it Suzaku} \citep{Mit07}, obtained in a different epoch of the {\it
XMM-Newton} observation. {\it Suzaku} carries (then available) three
CCD cameras called the X-ray Imaging Spectrometers, XIS-0, XIS-3
(Front-side Illuminated XIS; XIS-FI), and XIS-1 (Back-side Illuminated
XIS; XIS-BI), and a collimated-type instrument called the hard X-ray
detector (HXD), which consists of Si PIN photo-diodes and GSO
scintillation counters. The XISs and HXD-PIN covers the 0.2--12 keV
and 10--70 keV, respectively. Since AGNs are time variable, the
simultaneous coverage is critical for studying the continuum shape
over the broad band, in particular to accurately constrain the Compton
reflection component. Moreover, the {\it Suzaku} exposure is quite
deep (a net exposure of $\sim$ 100 ksec), and thus provides us with
the best quality dataset so far obtained from this source.

The organization of our paper is as follows. First, we describe our
{\it Suzaku} observation and data reduction in section~2. Next, we
present the analysis and results in section~3. The results of the {\it
Swift}/BAT data are also presented, whose spectrum averaged over 22
months is utilized to constrain the highest energy band up to 200
keV. We plot the spectral energy distribution of 4C 50.55 from the
radio to Gamma-ray bands to discuss the possible contribution of the
jet component. Finally, we discuss our results in comparison with the
previous studies of this target and other radio galaxies in section~4.
In all spectral analysis, we apply the Galactic absorption fixed at
\nhg $=1.0\times10^{22}$ cm$^{-2}$ \citep{Kal05}. The cosmological
parameters \cosmo\ = (71 km s$^{-1}$ Mpc$^{-1}$, 0.27, 0.73;
\citealt{Kom09}) are adopted in calculating the luminosities. The
errors attached to spectral parameters correspond to those at 90\%
confidence limits.

\section{Observations and Data Reduction}\label{obs}

4C 50.55 was observed with {\it Suzaku} from 2007 April 16 14:05 (UT)
to April 18 11:04 (UT) (observation ID 702027010), focused on the
nominal center position of the Hard X-ray Detector (HXD). The net
exposures after data screening, described below, are 85.0 ksec (XIS-0,
1), 80.0 ksec (XIS-3), 54.7 ksec (PIN). For the HXD, we use only data
collected with the PIN diodes, because we find that the signal from
the source in the GSO data is not significant over the systematic
error of $\approx 2\%$ in the non X-ray background (NXB)
\citep{Fuk08}. We use FTOOLS (heasoft version 6.6.2) to extract data,
and XSPEC version 11.3.2ag for spectral fitting.

\subsection{XIS Data Reduction}

To apply the latest calibration, we reprocess the unfiltered event
files of the XIS data according to the procedures written in {\it The
Suzaku Data Reduction Guide (ABC Guide)}. We select the data where the
time since the South Atlantic Anomaly (T\_SAA) passage is longer than
436 sec, the elevation angle (ELV) is larger than 5$^\circ$, and the
dye-elevation angle (DYE-ELV) is larger than 20$^\circ$. The XIS
events are extracted from a circular region centered on
the source peak, and the background is taken from a source-free region 
with the same distance from the optical axis as the target. 
We generate RMF files with {\it xisrmfgen}, and ARF
files with {\it xissimarfgen} \citep{Ish07}. We examine the spectra of
the $^{55}$Fe calibration source (producing an Mn K$\alpha$ line at
5.895 keV) located on the corners of the XIS chips to check the
accuracy of the energy response. By fitting them with Gaussians, we
are able to reproduce the right central energy within the statistical
errors and line widths of $3.9 \ (< 19.2)$ eV (XIS-0), $0.0 \ (<
15.6)$ eV (XIS-1), and $0.0 \ (< 14.0)$ eV (XIS-3). This verifies that
both energy scale and energy resolution in the responses are well
reliable.

\subsection{HXD-PIN Data Reduction}

We also reprocess the unfiltered event files of the HXD data as
well. This include the time assignment (with {\it hxdtime}), gain
correction ({\it hxdpi}), and grade classification ({\it
hxdgrade}). We select data where the time after the SAA passage is
longer than 500 sec, a cutoff rigidity (COR) is larger than 6 GV, and
the elevation from the earth is larger than 5$^\circ$. We utilize the
``tuned'' NXB event files provided by the HXD team to produce the
background spectra, to which the cosmic X-ray background (CXB) based
on the formula by \citet{Gru99} is added. We use the HXD/PIN response
file ae\_hxd\_pinhxnome3\_20080129.rsp.

\section{Analysis and Results}\label{ana}

\subsection{{\it Swift}/BAT Data}\label{BAT}

\begin{figure}
\epsscale{1.0}
\rotatebox{-90}{
\includegraphics[scale=0.3]{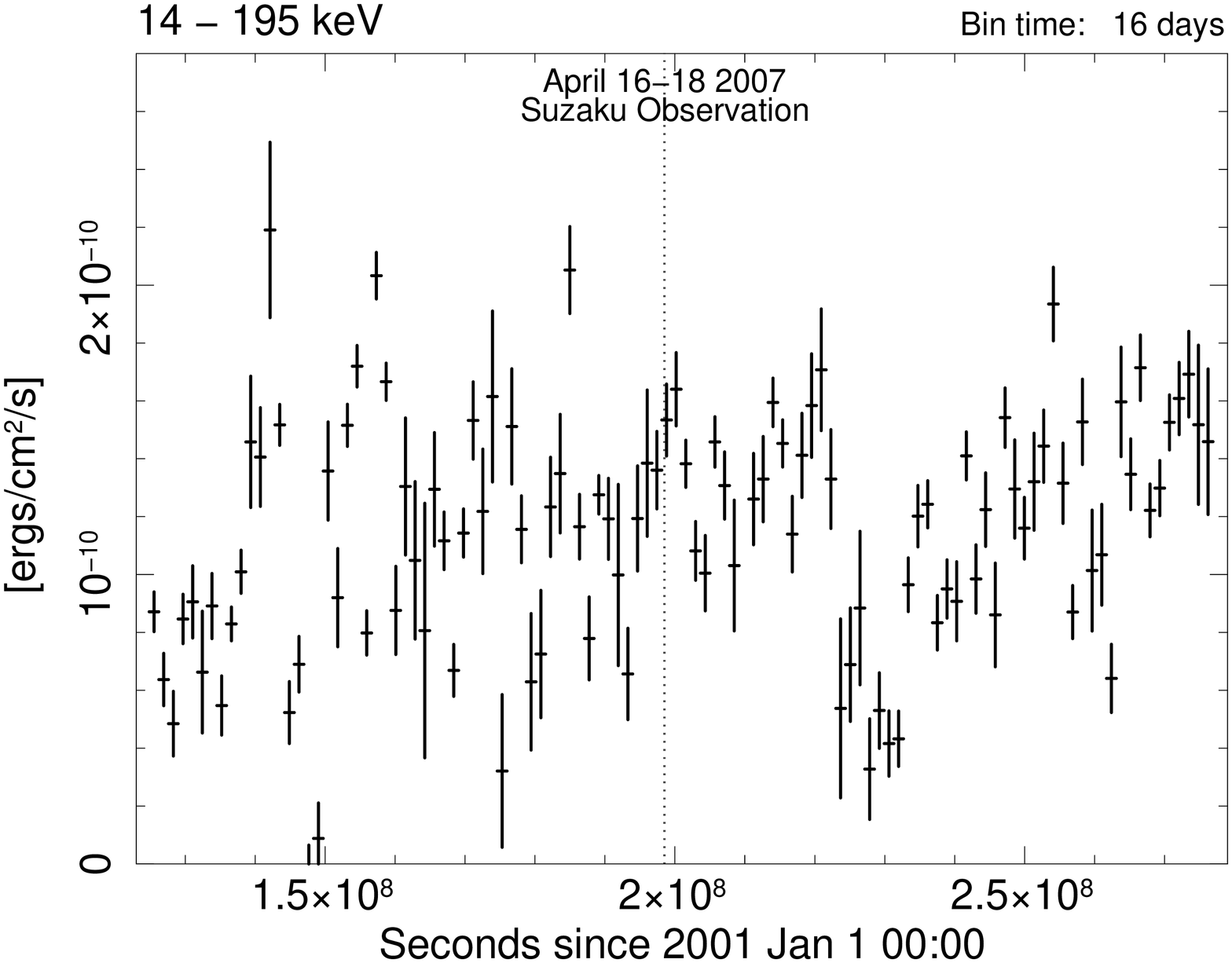}
}
\caption{
Long term light curve of 4C 50.55 in the 14--195 keV band obtained with
{\it Swift}/BAT with 16 days bin. The fluxes are converted from the
count rate by assuming a power law photon index of $\Gamma=1.7$. The
dotted lines indicate the period of the {\it Suzaku} observation
(from 2007 April 16 to April 18). \\
}
\label{curve_b}
\end{figure}

\begin{figure}
\epsscale{1.0}
\includegraphics[angle=270, scale=0.3]{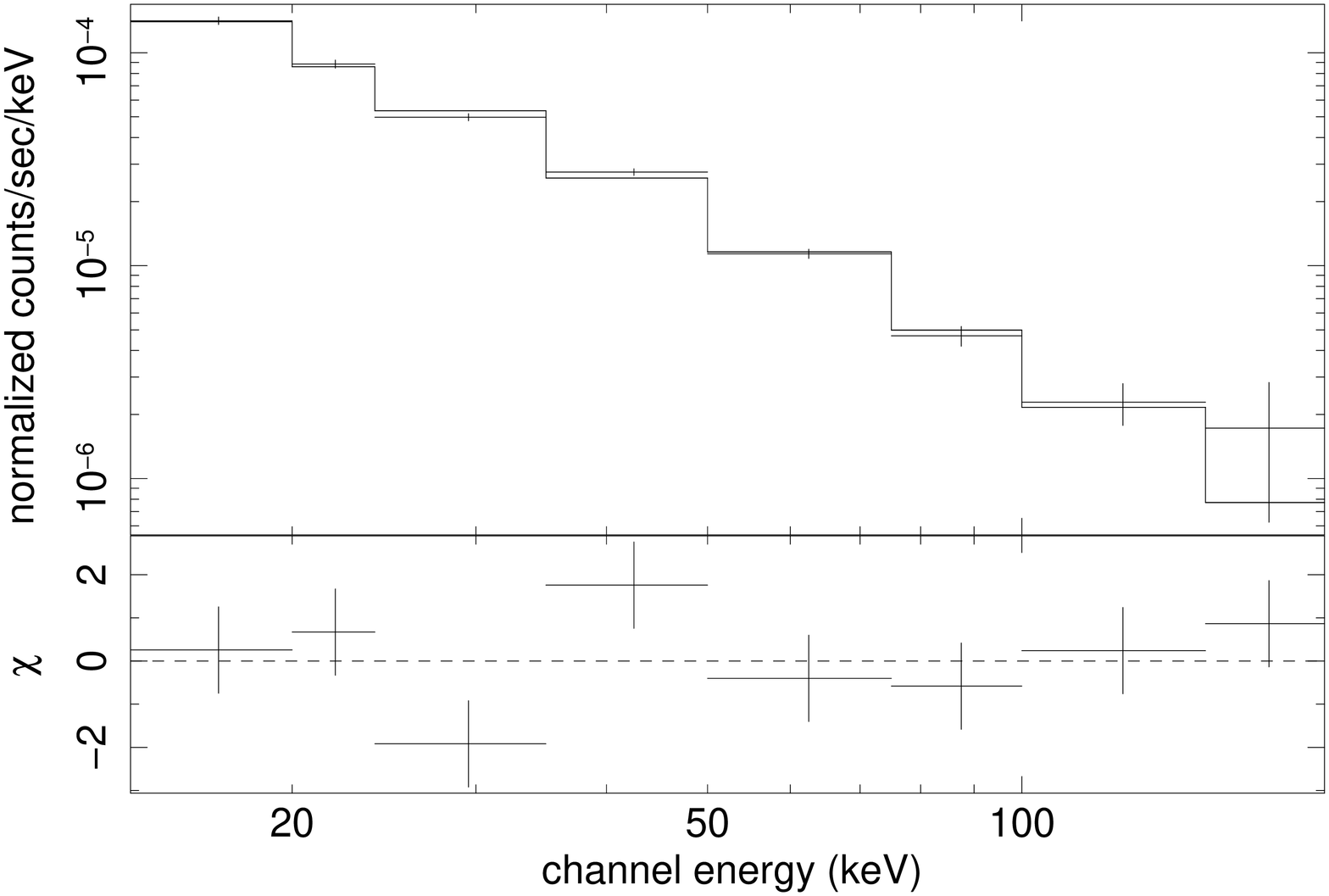}
\caption{ 
The time-averaged {\it Swift}/BAT spectrum of 4C 50.55 over 22 months.
The best-fit model is a pexriv model (see Table \ref{tab_s}). \\
}
\label{bat_spec}
\end{figure}

Figure~\ref{curve_b} shows the long term light curve of 4C 50.55 in
the 14--195 keV band obtained with {\it Swift}/BAT from 2004 December 15 to
2009 October 10. Each data point corresponds to the averaged flux for 16
days. Time variability is clearly noticed. The epoch of our {\it
Suzaku} observation is indicated by the dashed line in the figure. The
time-averaged BAT spectrum over the first 22 months covering the
14--195 keV band is plotted in Figure~\ref{bat_spec}. We find that it
can be fit with a single power law of $\Gamma=1.68\pm0.25$. The time
averaged 14--195 keV flux is $1.7 \times 10^{-10} \ {\rm ergs \
cm^{-2} \ s^{-1}}$ (based on the best-fit power law model), which
corresponds to a luminosity of $1.5 \times 10^{44} \ {\rm ergs \
s^{-1}}$ in the rest-frame 14.3--199 keV band. In the subsequent
subsections, we will apply more realistic spectral models to this
spectrum.

\subsection{{\it Suzaku} Light Curve}\label{Suzaku_curve}

\begin{figure}
\epsscale{0.7}
\includegraphics[angle=270, scale=0.3]{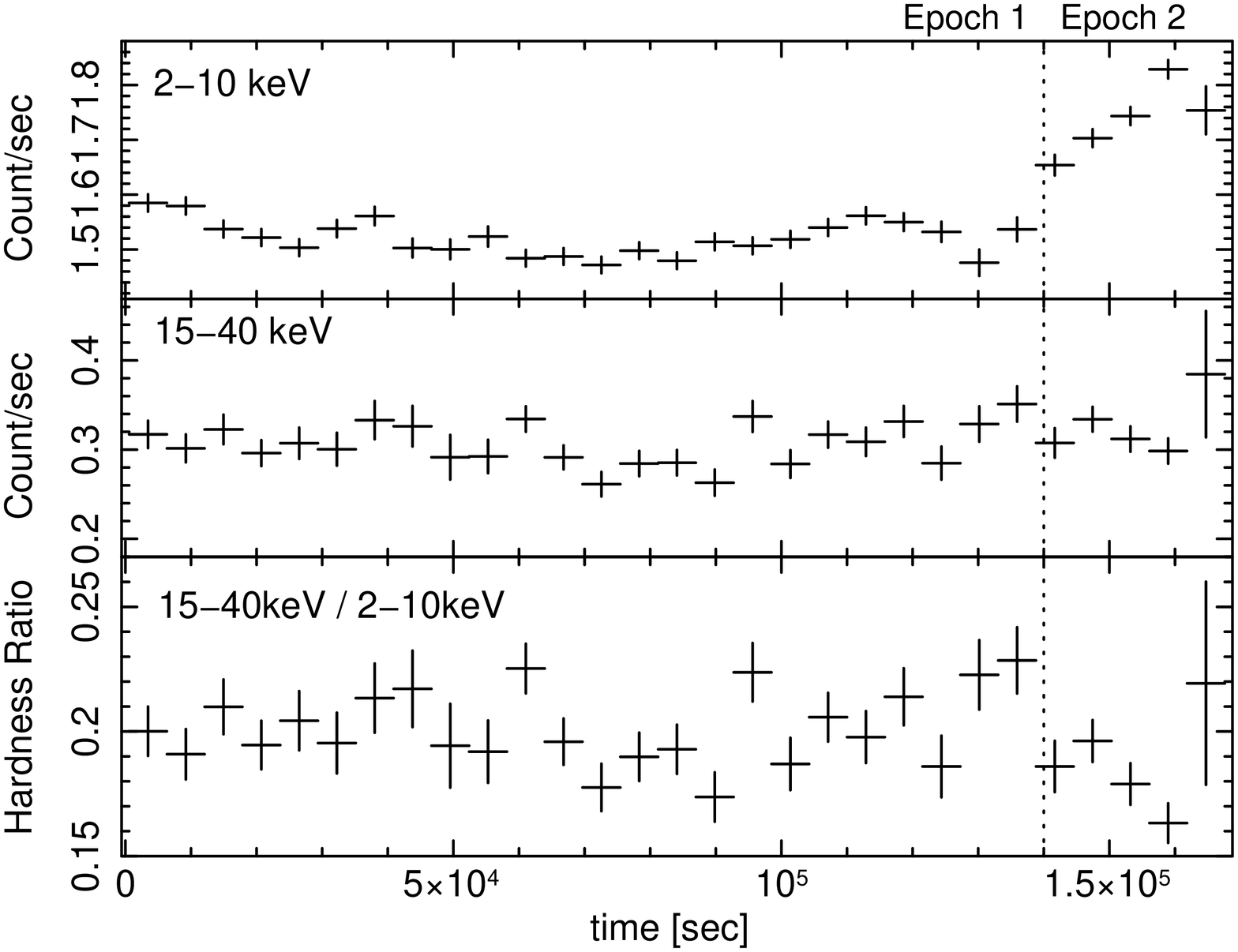}
\caption{
Light curves of 4C 50.55 obtained by the {\it Suzaku} observation, 
with a 5760 sec bin. The start time corresponds to 2007 April 16. 
The dotted line shows the border between epochs 1 and 2.
({\it upper}) the light curve of XIS-FIs in the 2--10 keV band.
({\it middle}) that of PIN in the 15 -- 40 keV band.
({\it lower}) the hardness ratio between them (15--40 keV / 2--10 keV). \\
}
\label{curve_s}
\end{figure}

We make the {\it Suzaku} light curves of 4C 50.55 with a 5760 sec bin,
the orbital period of the satellite, to remove any possible modulation
related to the orbital condition. Figure~\ref{curve_s} shows the X-ray
light curves obtained with the XIS-FIs (2--10 keV, {\it upper}) and
with the HXD/PIN (15--40 keV, {\it middle}), and their hardness ratio
(PIN/XIS, {\it lower}). The zero point of time corresponds to the
start time of the {\it Suzaku} observation. The XIS light curve in the
2--10 keV band indicates a rapid flux increase by $\sim 20\%$ in the
last 20 ksec exposure. The time scale of this variability is $\sim
10^4$ sec, indicating that the emission region is within $\sim 30
r_{\rm g}$ ($r_{\rm g} \equiv \frac{GM}{c^2}$ is the gravitational
radius) for a black hole mass of $M = 10^{7.8} \solarmass$ (see
section~\ref{differ_SED}). By contrast, the light curve in the 15--40
keV band does not show evidence for significant variability above the
statistical errors in the same epoch. Thus, there is a hint for
decrease in the hardness ratio between the PIN and XIS, although its
significance is marginal. For the following spectral analysis, we
separate the observation period into two, epoch~1 (0--1.4$\times 10^5$
sec) and epoch~2 (1.4$\times 10^5$--1.7$\times 10^5$ sec).

\subsection{Spectral Analysis with Phenomenological Models}\label{Suzaku_spec}

\subsubsection{Individual Fit to the {\it Suzaku} and {\it Swift} Data}

We analyze the {\it Suzaku} spectra of epoch~1 and epoch~2 separately,
by performing simultaneous fit to those of the XIS-FIs in the 1--9 keV
band, the XIS-BI in the 1--8 keV band, and the HXD/PIN in the 12--60
keV band. Given the fact that the spectrum of 4C 50.55 is subject to a
heavy Galactic absorption, we do not utilize the XIS-BI data below 1
keV to avoid any possible uncertainties in the response. We find that
the effects of pile-up are significant above 9.0 keV for the XIS
data. Further, we discard the XIS data of the 1.7--1.9 keV band
because of calibration uncertainties associated with the instrumental
Si-K edge. In the simultaneous fit, the flux normalizations for the
XIS-FIs and XIS-BI are set free, while that of the PIN is fixed at
1.18 relative to that of the XIS-FIs \citep{Mae08}.

We firstly apply a single power law model (modified by the Galactic
absorption fixed at \nhg $ = 1.0 \times 10^{22} \ {\rm cm^{-2}}$) to
the XIS and PIN data in epoch~1, covering the 1--60 keV band. The fit
is found to be far from acceptable ($\chi^2 / {\rm dof} = 13035/810$)
and results in a very flat slope, $\Gamma \approx 1.2$. This is
because the spectral shape below 10 keV is much harder than that above
10 keV. Next we apply a cut-off power law model, in the form of
$E^{-{\rm \Gamma}} \times {\rm exp}(-E/E_{\rm cut})$, over which a
narrow Gaussian is added to represent an iron-K emission line at 6.4
keV. We obtain ${\rm \Gamma} \approx 0.83$ and $E_{\rm cut} \approx
14$ keV with $\chi^2 /{\rm dof} = 4616/808$. There remain strong
absorption features around 1 keV suggesting intrinsic absorption,
however. When we add another absorption at the source redshift
($z=0.02$), the fit becomes much better, yielding $\Gamma
=1.39\pm0.02$, $E_{\rm cut} = 45\pm2$ keV, and $N_{\rm H} =
(0.67\pm0.01) \times 10^{22} \ {\rm cm^{-2}}$ with $\chi^2/{\rm dof} =
997/807$. Finally, similar to the analysis done by \citet{Mol07}, we
consider double absorber model with two different column densities,
\nhone\ and \nhtwo, whose covering fraction is $f$ and $(1-f)$,
respectively. The fit is significantly improved ($\chi^2/{\rm dof} =
940/805$) with $\Gamma = 1.61^{+0.03}_{-0/05}$, 
$E_{\rm cut} = 140^{+46}_{-20}$ keV,
\nhone$ \approx 8.3\times 10^{22} \ {\rm cm^{-2}}$, 
\nhtwo$ \approx 0.73 \times 10^{22} \ {\rm cm^{-2}}$, 
and $f= 0.19$. The photon index
becomes a reasonable value for AGNs. Note that our ``double absorber''
model has three free parameters for the absorber, while the ``double
partial covering'' model ({\bf ``pcf*pcf*''} in the XSPEC terminology)
adopted by \citet{Mol07} has four. Since the latter model does not
give a significant improvement for our data ($\Delta \chi^2 \approx 1$), we
adopt the former as a base continuum model for the following analysis.

The high quality {\it Suzaku} spectra are quite useful to constrain
the reflection component, which is indicated by the presence of the
iron-K emission line. Thus, we include it by utilizing our
modified version of ``pexriv'' reflection code \citep{Mag95} that
assumes a cutoff power law continuum and contains a self-consistent
fluorescence iron-K line calculated according to the same algorithm as
described in \citet{Zyc99}. The additional free parameter is the
relative reflection strength, $R (\equiv \Omega / 2\pi)$, while we fix
the ionization parameter, temperature, and inclination angle at
0, $10^5$ K, and 35 degrees \citep{Mol07}, respectively. The
solar Fe abundance by \citet{and89} (Fe/H = $4.68\times10^{-5}$) is
assumed. Considering that the reflection most likely occurs in the
accretion disk, we smear both reflected continuum and iron-K emission
line by the {\bf ``diskline''} kernel \citep{Fab89}. The innermost
radius $r_{\rm in}$ is set as a free parameter by assuming an
emissivity law of $r^{-3}$ for a fixed outer radius $r_{\rm out} =
10^5 r_{\rm g}$. When constraining $r_{\rm in}$, we utilize only the
XIS spectra around the iron-K band (3--9 keV for the XIS-FIs and 3--8
keV for the XIS-BI) and fix all the other parameters except for the
normalization. Thus, the spectral fit is performed by iteration; after
$r_{\rm in}$ is determined from the XIS-only fit, it is then fixed
when finally determining the continuum parameters in the XIS+PIN fit.

\begin{figure*}
\epsscale{1.0}
\includegraphics[angle=270, scale=0.3]{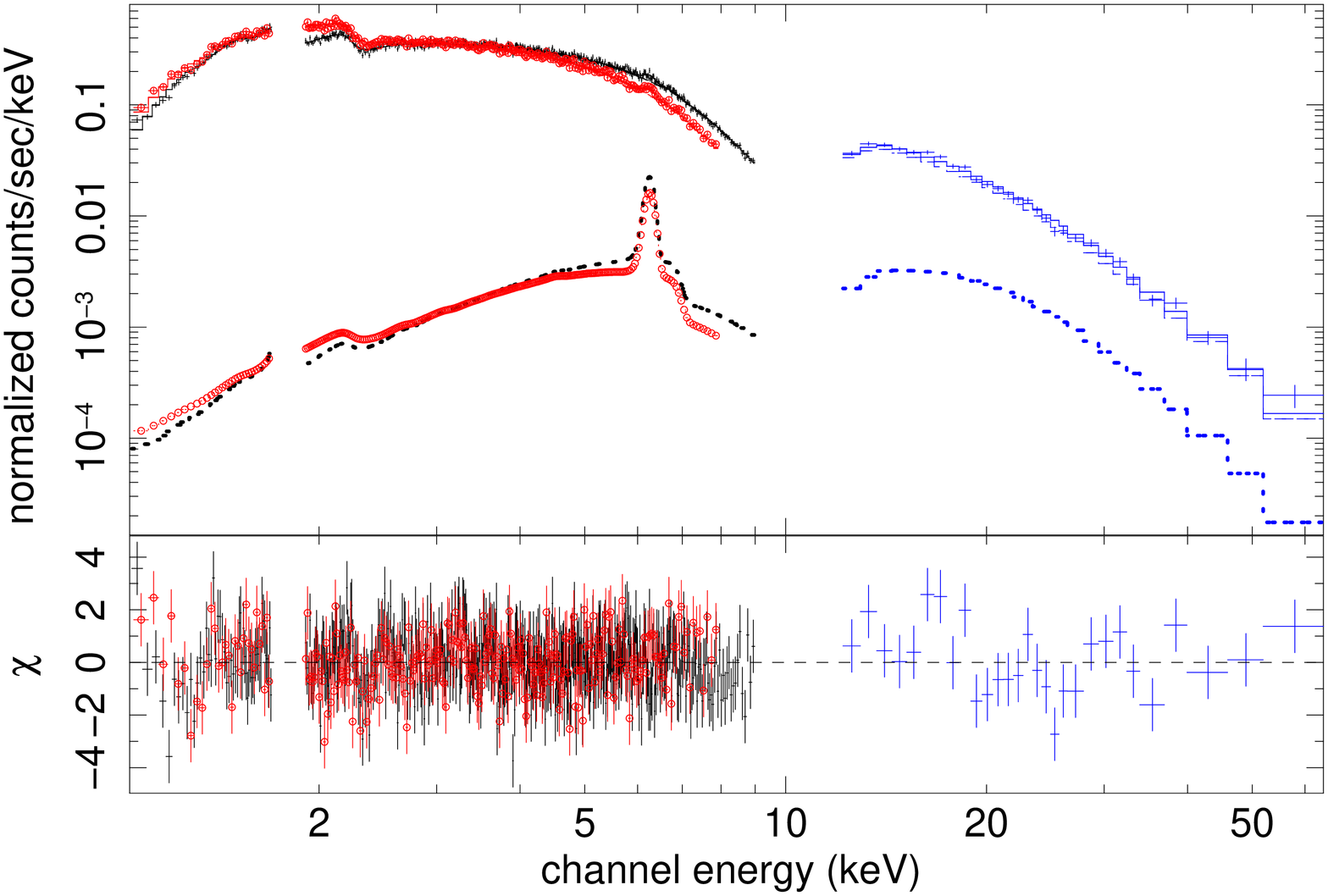}
\includegraphics[angle=270, scale=0.3]{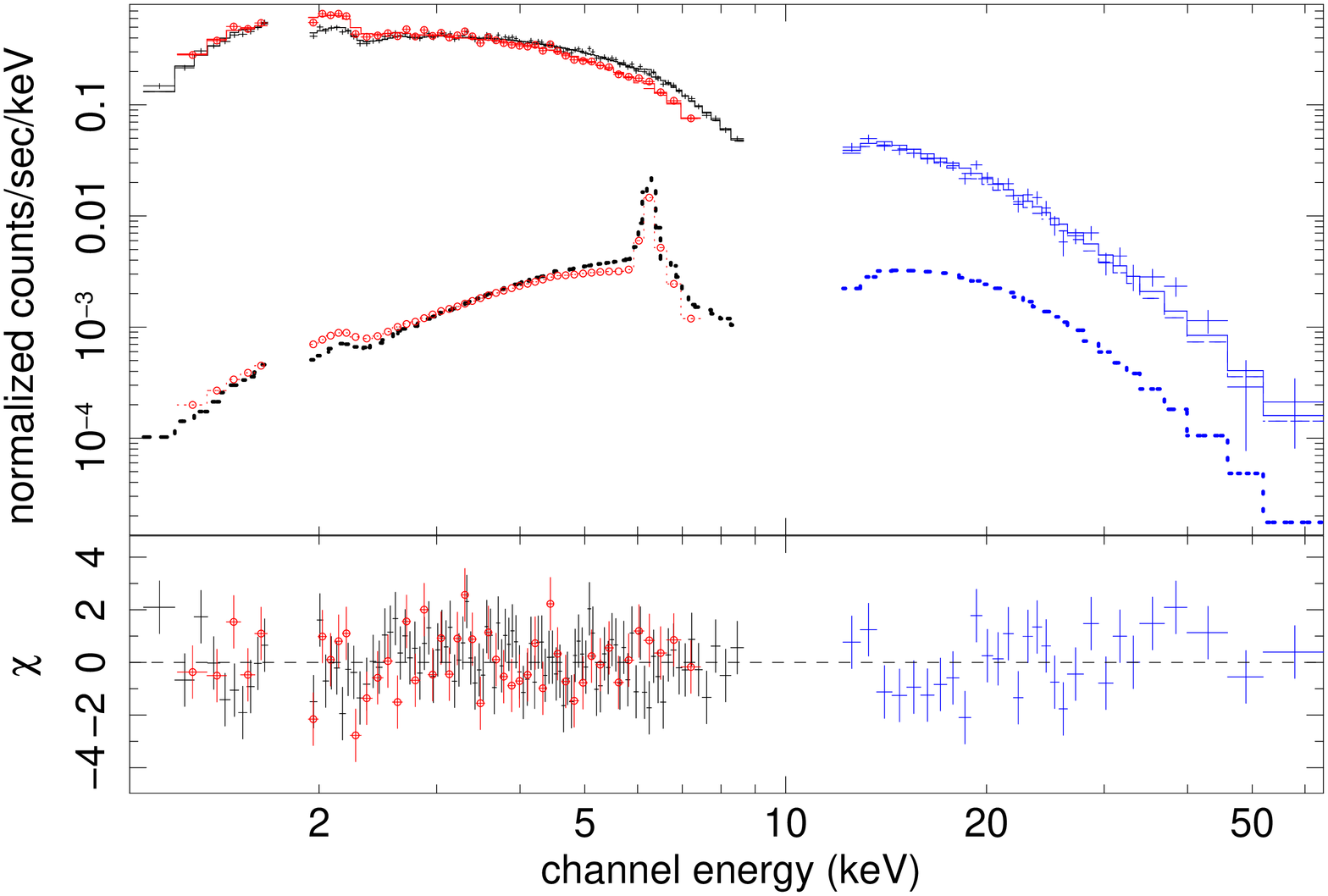}
\caption{ 
(a) ({\it left}) The folded spectra of 4C 50.55 in epoch~1 obtained
with the XIS-FIs in the 1--9 keV band (black), the XIS-BI in the 1--8
keV band (red, open circle), and the PIN in the 12--60 keV band
(blue). The solid curve represents the best-fit model (see Table
\ref{tab_s}). The dotted curve represents the reflection component.
(b) ({\it right}) those in epoch~2. \\}
\label{F_spec}
\end{figure*}

\begin{figure}
\epsscale{1.0}
\includegraphics[angle=270, scale=0.3]{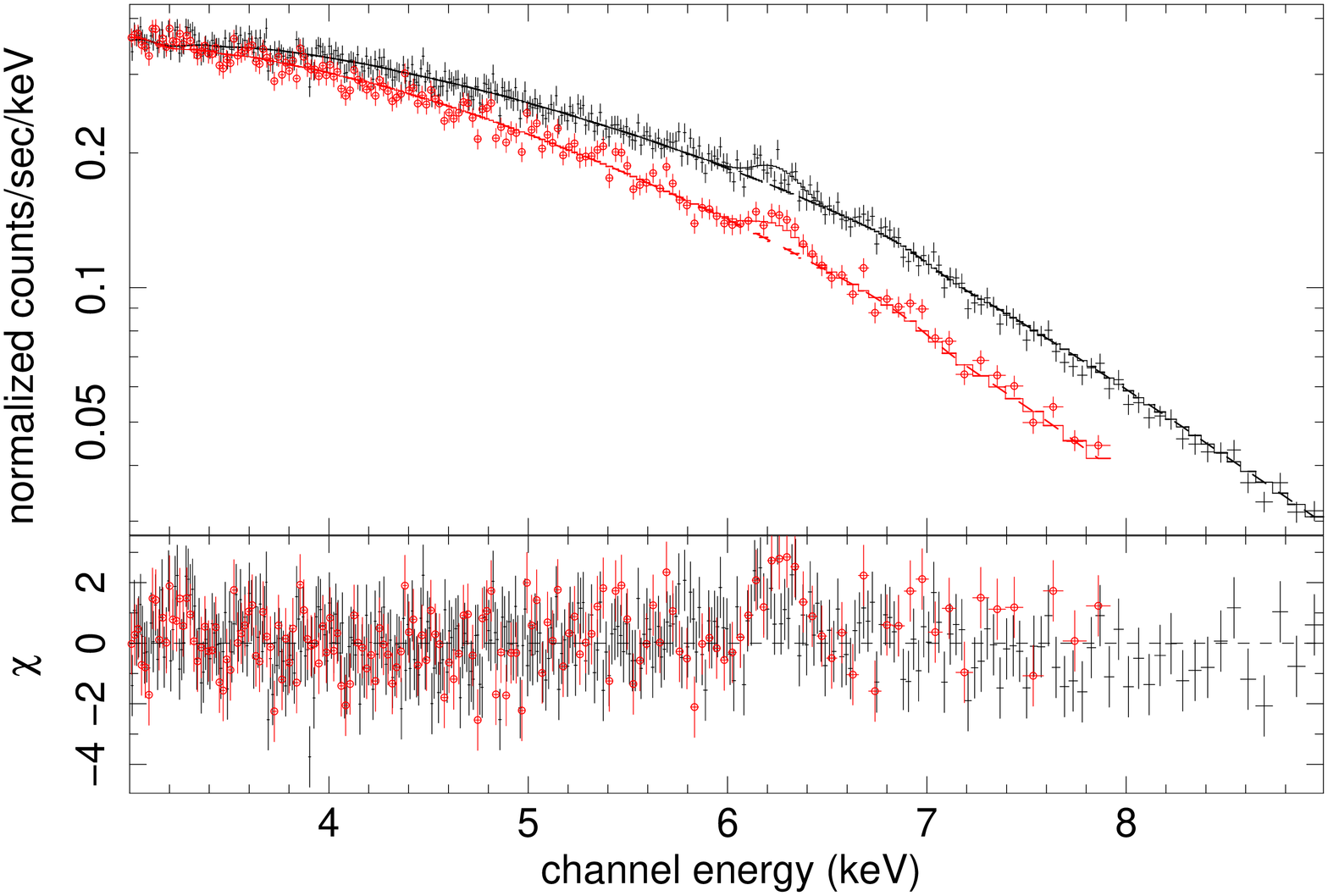}
\caption{
({\it upper}) The spectra of 4C 50.55 obtained with the XIS-FIs
(black) and XIS-BI (red, open circle) in epoch~1. The best-fit model (see
Table~\ref{tab_s}) is plotted by the solid curve.
The dashed curve indicates the model from which the iron-K emission line 
is excluded.  
({\it lower}) The residuals of the fit in units of $\chi$
for the model without the iron-K line.  \\ }
\label{Fe}
\end{figure}

The results of the spectral fit to the individual {\it Suzaku} spectra
in epochs 1 and 2 are summarized in
Table~\ref{tab_s}. Figure~\ref{F_spec} shows the XIS+PIN spectra
folded by the energy responses, over which the best-fit models are
plotted, with residuals in the lower panel. The expanded figure of the
XIS spectra between 3--9 keV in epoch~1 is plotted in
Figure~\ref{Fe}. To emphasize the iron-K line feature, the residuals
when the line is excluded from the model are shown in the lower panel
of this figure. From epoch~1, we obtain $\Gamma = 1.61\pm0.05$,
$E_{\rm cut} = 80^{+36}_{-19}$ keV, and $R = 0.18\pm0.04$. 
The innermost radius is constrained to be $r_{\rm in} = 720 r_{\rm g}
(> 340 r_{\rm g})$ from the XIS data. For the analysis of the epoch 2
spectra, we assume the same parameters of the reflection component
(including its absolute flux) as those found from the epoch~1 data,
since it is very unlikely that it varied on such a short time scale of
$< 10^4$ sec. The parameters of the absorption are fixed to the
epoch~1 values as well. Thus, only free parameters are $\Gamma$,
$E_{\rm cut}$, and the normalization. Finally, we also perform
spectral fit to the time-averaged {\it Swift}/BAT spectrum by adopting
the same model. The reflection strength is fixed at $R=0.18$ referring
to the epoch~1 result. The best-fit model is over-plotted in
Figure~\ref{bat_spec}, whose parameters are summarized in Table
\ref{tab_s}.

\begin{deluxetable}{cccc}
\tabletypesize{\footnotesize}
\tablecaption{Best-Fit Parameters of Cut-off Power Law Model to the Individual {\it Suzaku} and {\it Swift} Data\label{tab_s}}
\tablewidth{0pt}
\tablehead{\colhead{Parameters} & \colhead{Epoch 1} & \colhead{Epoch 2} & \colhead{\it Swift/BAT}}
\startdata
$^{\rm (a)}f_{2-10 \ {\rm keV}}$ & $7.4 \times 10^{-11}$ & $ 8.5 \times 10^{-11} $ & -- \\
$^{\rm (a)}f_{10-60 \ {\rm keV}}$ & $1.6 \times 10^{-10}$ & $ 1.6 \times 10^{-10} $ & -- \\
$^{\rm (a)}f_{14-195 \ {\rm keV}}$ & -- & -- & $1.7 \times 10^{-10}$ \\
\hline
$^{\rm (b)}$\nhg & 1.0$^\S$ &  1.0$^\S$ & --  \\
$\Gamma $  & $1.61 \pm 0.05$ & $1.64 \pm 0.03$ & $1.68 ^{+0.23}_{-0.25}$ \\
$E_{{\rm cut}}$ [keV]  & $80 ^{+36}_{-19}$ & $68 ^{+29}_{-16}$ & $127 ^{+340}_{-57}$ \\ 
$R (= \Omega / 2 \pi)$ &  $0.18 \pm 0.04$ & $*$  & 0.18$^\S$ \\
$^{\rm (c)}f$  & $0.19 ^{+0.03}_{-0.04}$ & 0.19$^\S$ & --  \\
$^{\rm (d)}$\nhone & $7.6 ^{+1.5}_{-1.4}$  & 7.6$^\S$ & -- \\
$^{\rm (d)}$\nhtwo & $0.73 \pm 0.03$  & 0.73$^\S$ & -- \\ 
$EW$ [eV]  & 22 & 19 & --     \\  
$r_{\rm in} [r_{\rm g}]$ & 720$\S$ $( > 340)^\dag$ & 720$^\S$ & 720$^\S$  \\
\hline
$\chi ^2 /$ dof & $916.0 / 805$ & $190.8 / 175$ &  $8.594 / 5$           \\
\enddata
\tablecomments {Errors are $ 90 \% $ confidence level for a single parameter. \\
$^{\rm (a)}$ Observed fluxes in the 2--10 keV, 10--60 keV, and 14--195 keV bands, in units of ${\rm ergs \ cm^{-2} \ s^{-1}}$.\\
$^{\rm (b)}$ Galactic absorption column density in units of $10^{22} \ {\rm cm^{-2}}$. \\
$^{\rm (c)}$ Covering fraction\\
$^{\rm (d)}$ Intrinsic absorption column density at the source redshift in units of $10^{22} \ {\rm cm^{-2}}$. \\
$^\S$ Parameters fixed at these values. \\
$*$ We assume the same reflection component as that determined in epoch 1.\\
$^\dag$ We constrain the inner radius only from the 3---9 keV XIS
spectra in epoch 1, which is then fixed at the best-fit when determining
the continuum parameters. 
\\ }
\end{deluxetable}

\subsubsection{Simultaneous Fit to the {\it Suzaku} and {\it Swift} Data}

From the above analysis, we find no significant differences in the
spectral parameters (except for the normalization) within the
statistical errors between the {\it Suzaku} epoch~1, epoch~2, and {\it
Swift}/BAT data, although there is a hint that the spectrum became
slightly softer in epoch~2. Thus, to best constrain the continuum
parameters, in particular the cutoff energy, we study the {\it Suzaku}
spectra (either of the two epochs) in the 1--60 keV band and {\it
Swift} spectrum in the 14--195 keV band simultaneously, in all
following analysis. The flux normalization between the {\it Suzaku}
(XIS-FIs) and BAT spectra are set free, to take into account the time
variability. In analyzing the spectra of epoch 2, we always fix the
reflection component to that determined from the epoch~1 data.

Table \ref{tab_rfcut} summarizes the results using the same phenomenological
model (cutoff power law) as adopted in section \ref{Suzaku_spec}. For
epoch~2, we consider two extreme cases as the cause of the spectral
variability from epoch~1 that (1) only the continuum changed without
change of the absorber and that (2) only the absorber changed with the
same continuum except for its normalization. We obtain similarly good
fits for the two cases, and thus both possibilities are plausible from
the spectral analysis. In reality, however, it may be difficult to
explain such a short time ($<10^4$ sec) variability by the absorber
alone. If the absorber makes Kepler motion at $\sim$1000 $r_{\rm g}$,
a typical location of the broad line region in AGNs, it moves only
$\sim$ $r_{\rm g}$ in $10^4$ sec, by assuming the black hole mass of
$10^{7.8}$ \solarmass. Thus, unless the emitting region is extremely
small (like $<$ several $r_{\rm g}$), it is unlikely that crossing
blobs in the line of sight can cause the large variability as observed.

\begin{deluxetable}{cccc}
\tabletypesize{\footnotesize}
\tablecaption{Best-Fit Parameters of Cut-off Power Law Model to the Combined {\it Suzaku} and {\it Swift} Data\label{tab_rfcut}}
\tablewidth{0pt}
\tablehead{\colhead{Parameters} & \colhead{Epoch 1} & \colhead{Epoch 2 (1)} & \colhead{Epoch 2 (2)}}
\startdata
$^{\rm (a)}L_{2 - 10 \ {\rm keV}}$ & 8.1 & 9.3 & 9.0 \\
\hline
$^{\rm (b)}$\nhg & 1.0$^\S$ & 1.0$^\S$ & 1.0$^\S$  \\ 
$\Gamma $ & $1.65 \pm 0.04$ & $1.69 \pm 0.02$ & 1.65$^\S$  \\
$E_{\rm cut}$ [keV] & $105 ^{+28}_{-19}$ & $127 ^{+25}_{-23}$ & 105$^\S$ \\ 
$R (= \Omega / 2 \pi)$ & $0.17 \pm 0.04$ & $*$ & $*$   \\ 
$EW$ [eV] (Fe K$\alpha$ line)  & 21 & 18 & 18 \\
$^{\rm (c)}f$ & $0.21 \pm 0.03$ & 0.21$^\S$ & $0.40 ^{+0.26}_{-0.15}$ \\
$^{\rm (d)}$\nhone & $7.9 ^{+1.4}_{-1.3}$ & 7.9$^\S$ & $2.9 ^{+1.6}_{-1.0}$  \\
$^{\rm (d)}$\nhtwo & $0.75 \pm 0.03$ & 0.75$^\S$ & $0.49 ^{+0.17}_{-0.33}$ \\
\hline
$\chi ^2 /$ dof  & $926.9 / 812$ & $201.1 / 182$ & $192.7 / 181$           \\
\enddata
\tablecomments {Errors are $ 90 \% $ confidence level for a single parameter. \\
$^{\rm (a)}$ Intrinsic luminosity in the 2--10 keV band corrected for both Galactic and intrinsic absorptions in units of $ 10^{43} \ {\rm ergs \ s^{-1}}$. \\
$^{\rm (b)}$ Galactic absorption column density in units of $10^{22} \ {\rm cm^{-2}}$. \\
$^{\rm (c)}$ Covering fraction.\\ 
$^{\rm (d)}$ Intrinsic absorption column density in units of $10^{22} \ {\rm cm^{-2}}$. \\
$^\S$ Parameters fixed at these values. \\
$*$ We assume the same reflection component as that determined in epoch 1.\\
}
\end{deluxetable}

\subsection{Spectral Analysis with Comptonization Model}\label{thComp}

In this subsection, we analyze the spectra of 4C 50.55 with a
physically motivated model instead of the phenomenological ``cutoff
power law'' model, which is a mathematical approximation of the X-ray
spectra of AGNs. Such analysis of AGN spectra has been very limited so
far, since it requires high quality broad band data. As we will
discuss in section~\ref{differ_SED}, the contribution from the jet
components is very small in the X-ray band. Hence, we consider that
the origin of the continuum emission is predominantly thermal
Comptonization of soft (ultra-violet) photons off hot electrons in the
corona located above the accretion disk. Accordingly, we adopt a
thermal Comptonization model, {\bf thComp} \citep{Zyc99}, for the
primary continuum. It has two free parameters, the slope $\Gamma$ and
electron temperature $kT_{\rm e}$. The electron scattering optical
depth $\tau_{\rm e}$ is related to $T_{e}$ and $\Gamma$ by the
following formula \citep{Sun80}:
\begin{eqnarray}
\tau_{\rm e} = \sqrt{2.25 + \frac{3}{(T_{\rm e}/511 \ {\rm keV})[(\Gamma + 0.5)^2 -2.25]}}-1.5.
\end{eqnarray}
For seed photons, we assume a multicolor disk component with the
innermost temperature of 0.01 keV. This choice is not important to
constrain the Comptonization parameters. As described in the previous
subsection, the reflection component and fluorescence iron-K line are
self-consistently included, which are blurred by the ``diskline''
profile with the same parameters as obtained above ($r_{\rm in} =720
r_{\rm g}$).

\begin{figure*}
\epsscale{1.0}
\includegraphics[angle=270, scale=0.3]{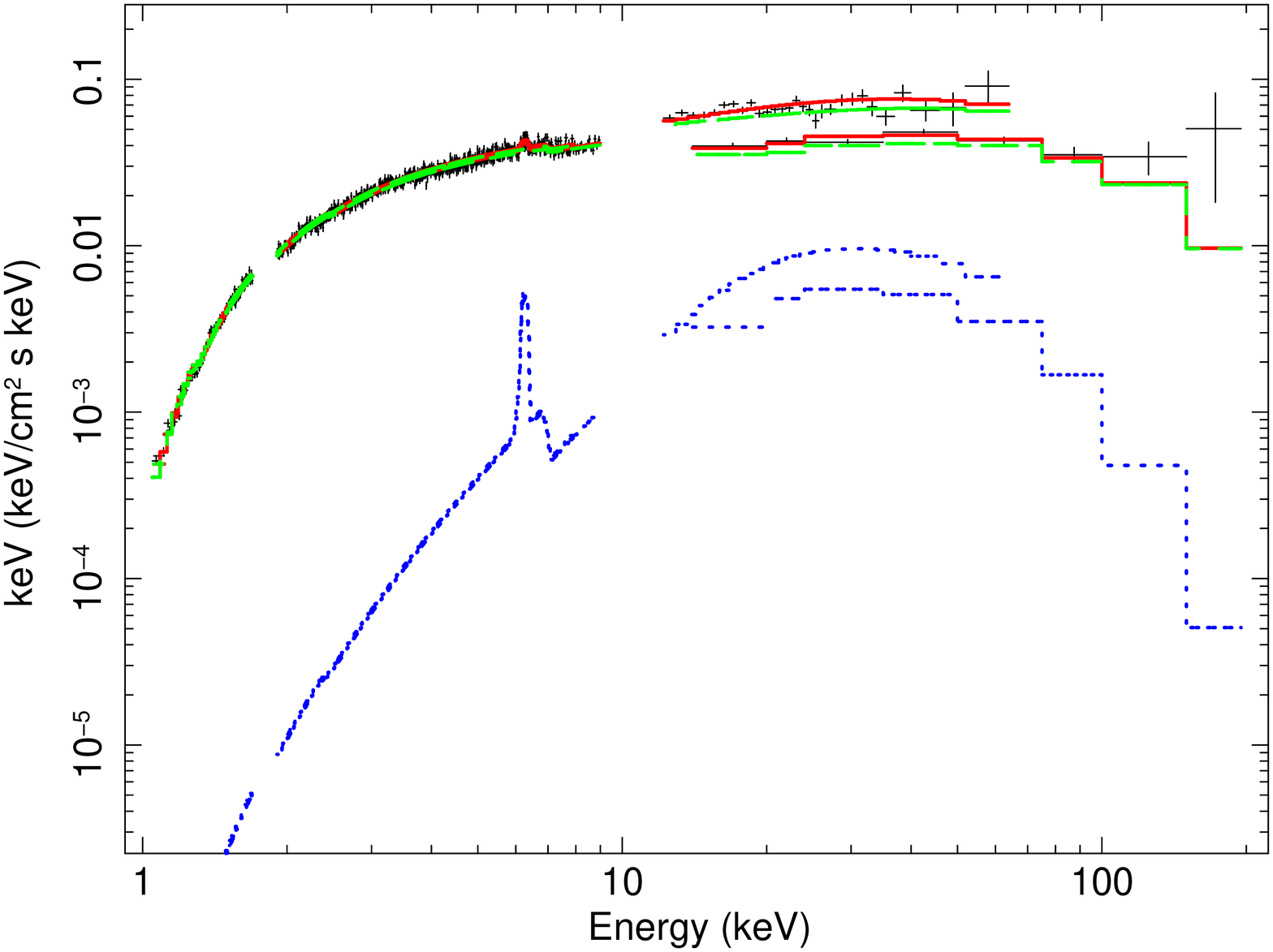}
\includegraphics[angle=270, scale=0.3]{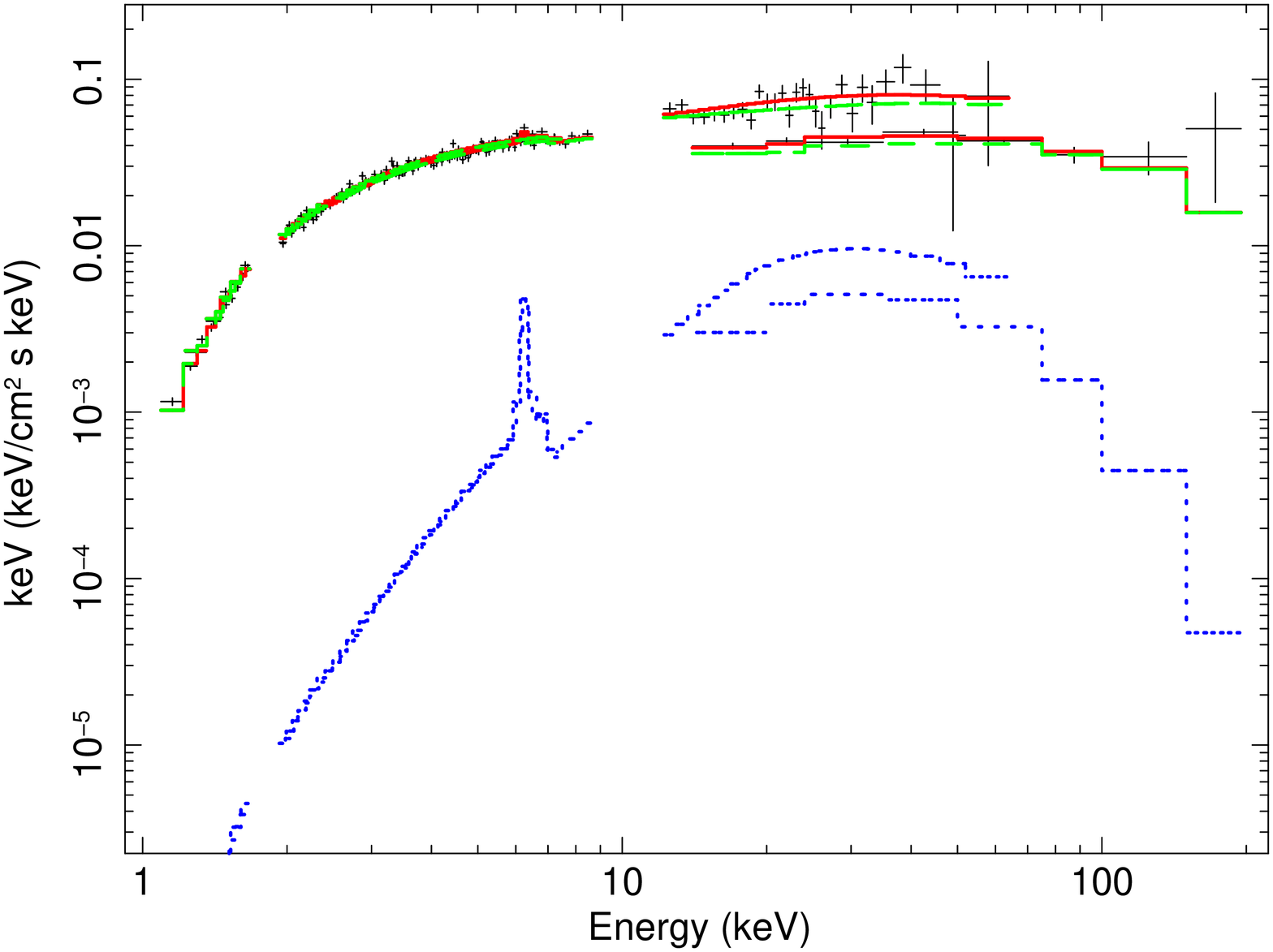}
\caption{ 
(a) ({\it left}) The unfolded spectra of 4C 50.55 determined from the
{\it Suzaku} data in epoch~1 and from the {\it Swift}/BAT data, based
on a thermal Comptonization model.
The crosses (black) represent the data points, dashed curve (blue) 
the transmitted component, dashed curve (magenta) the reflection component, 
and solid curve (red) the total. The best-fit parameters are given in Table~\ref{tab_thComp}.
(b) ({\it right}) those in epoch~2.
}
\label{F_thComp}
\end{figure*}

We perform simultaneous fit to the {\it Suzaku} (epoch~1 or 2) and
{\it Swift}/BAT spectra using this model. The best-fit parameters are
summarized in Table \ref{tab_thComp}. We obtain the electron
temperature of $\approx$30 keV with an optical depth of $\approx$
3. Again, the reflection component in epoch 2 are fixed at the same
one determined from epoch~1. We find that the continuum parameters are
consistent each other between epochs 1 and 2 within the statistical
errors at the 90\% confidence level. Figure~\ref{F_thComp} shows the
results in the two epochs, in the form of unfolded spectra (i.e.,,
those corrected for the effective area of the instruments) in units of
$E I(E)$, where $E$ is photon energy and $I(E)$ the energy flux.

\begin{deluxetable}{ccc}
\tabletypesize{\footnotesize}
\tablecaption{Best-Fit Parameters of Comptonization Model to the Combined {\it Suzaku} and {\it Swift} Data\label{tab_thComp}}
\tablewidth{0pt}
\tablehead{\colhead{Parameters} & \colhead{Epoch 1 \& BAT} & \colhead{Epoch 2 \& BAT}}
\startdata
$^{\rm (a)}L_{2 - 10 \ {\rm keV}}$ & 8.4 & 9.6 \\
\hline
$^{\rm (b)}$\nhg & 1.0$^\S$ & 1.0$^\S$   \\ 
$\Gamma $ & $1.78 \pm 0.02$ & $1.81 \pm 0.01$  \\
$kT_{\rm e}$ [keV] & $31 ^{+10}_{-6}$ & $46 ^{+37}_{-13}$ \\
$^{\rm (c)}$( \ \ $\tau_{\rm e}$ & $2.8^{+0.4}_{-0.5}$ &  $2.1^{+0.5}_{-0.8}$ \ \ ) \\ 
$R (= \Omega / 2 \pi)$ & $0.16 \pm 0.04$ & $*$    \\ 
$EW$ [eV] (Fe K$\alpha$ line)  & 19 & 17 \\
$^{\rm (d)}f$ & $0.27 \pm 0.02$ & 0.27$^\S$  \\
$^{\rm (e)}$\nhone & $8.6 ^{+1.1}_{-1.0}$ & 8.6$^\S$  \\
$^{\rm (e)}$\nhtwo & $0.80 \pm 0.03$ & 0.80$^\S$  \\
\hline
$\chi ^2 /$ dof  & $942.5 / 812$ & $200.8 / 182$           \\
\enddata
\tablecomments {Errors are $ 90 \% $ confidence level for a single parameter. \\
$^{\rm (a)}$ Intrinsic luminosity in the 2--10 keV band corrected for both Galactic and intrinsic absorptions in units of $ 10^{43} \ {\rm ergs \ s^{-1}}$. \\
$^{\rm (b)}$ Galactic absorption column density in units of $10^{22} \ {\rm cm^{-2}}$. \\
$^{\rm (c)}$ Electron-scattering optical depth calculated from the equation (1). \\
$^{\rm (d)}$ Covering fraction.\\
$^{\rm (e)}$ Intrinsic absorption column density in units of $10^{22} \ {\rm cm^{-2}}$. \\
$^\S$ Parameters fixed at these values. \\
$*$ We assume the same reflection component as that determined in epoch 1.\\
}
\end{deluxetable}

\subsection{Multi-Wavelengths Spectral Energy Distribution}\label{differ_SED}

\begin{figure}
\epsscale{0.7}
\rotatebox{-90}{
\plotone{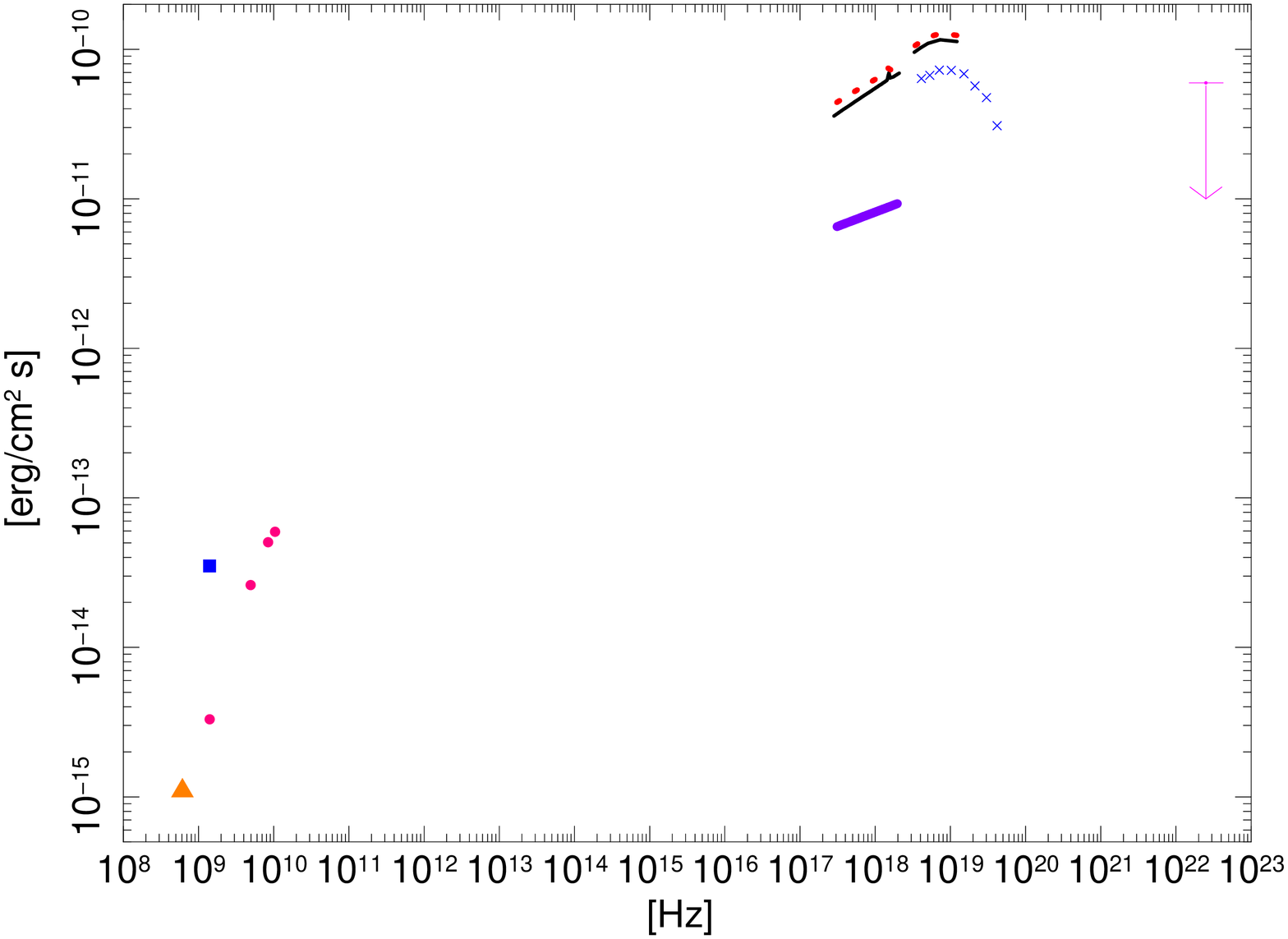}
}
\caption{
The SED of 4C 50.55. The solid (black) and dotted (red) curves
represent the {\it Suzaku} spectra in epochs 1 and 2, respectively,
while the thick (purple) one represents their difference spectrum between the
two epochs. The spectrum of {\it Swift}/BAT is shown by the blue
crosses.  The pink arrow represents the upper limit on the $\gamma$-ray
flux from the {\it Fermi} data \citep{Abd09}.  The radio data are
taken after \citet{Mol07}; the orange filled-triangle and reddish violet
filled-circles represent the core flux obtained from the GMRT
\citep{Pan06} and the VLA, respectively.  The blue filled-square
represent the total flux including both core and lobes.
}
\label{SED}
\end{figure}

We summarize in Figure~\ref{SED} the available multi-wavelengths
fluxes of 4C 50.55 to discuss its spectral energy distribution (SED)
from the radio to $\gamma$-ray bands. In the radio band, we utilize
those obtained by the VLA and the Giant Metrewave Radio Telescope
(GMRT) compiled by \citet{Mol07}. In the GeV $\gamma$-ray band, the
upper limit derived from the one-year {\it Fermi} data \citep{Abd09}
is indicated by the arrow. We plot the results of cutoff power law fit
to the {\it Suzaku} spectra in epochs 1 (black) and 2 (red) as well as
the time-averaged {\it Swift}/BAT data, correcting for both Galactic
and intrinsic absorptions. To examine the origin of the time
variability, we analyze the difference spectrum of XIS-FIs between
epochs 2 and 1. Fitting with a single power law with the same double
absorption as given in Table~\ref{tab_s}, we find $\Gamma \approx
1.8$. This result is also plotted in the figure (green curve).

Assuming that the global intrinsic SED of the jet emission in 4C~50.55
is similar to that of jet dominated-sources \citep{Kub98} with
correction for an estimated beaming factor ($\delta \sim 1$ for 4C~50.55
and $\delta \sim 10$ for blazars), the predicted X-ray emission is a factor
of $\sim 10^{1-4}$ less than the observed X-ray flux. We thus conclude
that X-ray emission due to a jet via synchrotron emission or inverse
Compton should be very small in the total X-ray emission of this
source. Based on the {\it Suzaku} results, the ratio of the core
luminosity at 5 GHz and that in the 2--10 keV band is estimated to be
log $R_{\rm X} = -3.6$. Thus, while 4C 50.55 should be classified as a
radio loud object \citep{Ter03}, its radio to X-ray power ratio is
much lower compared with typical blazars and more powerful radio
galaxies, like 3C~120 (log $R_{\rm X} = -2.1$) \citep{Kat07}. This
confirms the conclusion by \citet{Mol07}, who did not find any
evidence for an additional power law component representing the jet
emission in the {\it XMM-Newton} spectrum.

We note that the optical fluxes of 4C 50.55 are highly uncertain at
present due to the extinction (both Galactic and intrinsic ones),
which could largely affect the empirical estimate of the black hole
mass. In fact, the extinction-corrected AGN continuum flux at 5100
\AA\ adopted by \citet{Win10} falls far below an extrapolation from
the X-ray spectra toward lower energies. As a more simple approach,
assuming typical SEDs of AGNs \citep{Elv94,Gru10}, we roughly estimate
the 5100 \AA\ flux to be $(0.7-8) \times 10^{-10}$ erg
cm$^{-2}$ s$^{-1}$ from the BAT flux averaged for 22 months. Then,
from the luminosity at 5100 \AA, $(0.6-7) \times 10^{44}$ erg
s$^{-1}$, and the width of the H$_{\beta}$ line, 2320 km $ s^{-1}$
FWHM \citep{Win10}, we derive the black hole mass of 4C 50.55 to be
$\sim 10^{7.8\pm0.3}$ \solarmass, using the formula given by
\citet{Ves06}. This is about 10 times higher than the estimate
($10^{6.58 \pm 0.07}$ \solarmass) by \citealt{Win10}. Nevertheless,
the suggestion that the black hole in 4C 50.55 accretes with a high
Eddington ratio still holds. With a typical bolometric correction
factor ($L_{\rm bol}/L_{\rm 2-10 \; keV} = $30), we estimate the
fraction of Eddington luminosity of 4C 50.55 to be $L_{\rm bol}/L_{\rm
Edd} \sim 0.4$.

\section{Summary and Discussion}\label{discussion}

We have obtained the first simultaneous, broad band X-ray spectra of
4C~50.55 over the 1--60 keV band with {\it Suzaku}. These provide one
of the best quality high energy data so far obtained from BLRGs. From
the SED, we conclude that there is little contribution from jets in
the X-ray emission, which is thus dominated by Comptonization by hot
corona. We find that the overall continuum is represented by a cutoff
power law, or a thermal Comptonization model, with complex intrinsic
absorptions. We show that at least two absorber with different
covering fractions and column densities are necessary to model the
spectrum. The photon index and cutoff energy of 4C~50.55 are within
the distribution of previous observations of BLRGs \citep{Gra06},
although both are somewhat smaller ($\Gamma \approx 1.6$ and $E_{\rm
cut} \approx 100$ keV) compared with typical values found from Seyfert
galaxies ($\Gamma \approx $1.9 and $E_{\rm cut} \sim $200 keV,
\citealt{Dad08}). These findings on 4C~50.55 are fully consistent with
the results by \citet{Mol07} from the {\it XMM-Newton} and {\it
INTEGRAL} data observed on earlier epochs. We estimate that the corona
responsible for Comptonization is optically thick for scattering
$\tau_e \approx 3$ and is relatively cool $T_{\rm e} \approx 30$
keV. The optical depth is larger and the temperature is lower than
those obtained from Seyfert galaxies with thermal Comptonization
models \citep[e.g.,][]{Lub10}, as expected from the comparison in
$\Gamma$ and $E_{\rm cut}$.

From our observations, time variability of the X-ray flux on both long
($\simgt 10^6$ sec) and short ($\sim 10^4$ sec) time scales is
detected. The averaged 2--10 keV flux was the highest in our {\it
Suzaku} observation among those in the previous observations reported
in \citet{Mol07} by a factor of 1.3--2.5. The hard X-ray flux in the
17--100 keV band ($1.8\times10^{-10}$ erg cm$^{-2}$ s$^{-1}$) was also
higher by 1.4 than the averaged flux over 22-months obtained with {\it
Swift}/BAT ($1.3\times10^{-10}$ erg cm$^{-2}$ s$^{-1}$). The large
long-term variability in the hard X-ray above 10 keV suggests that it
is mainly produced by the intrinsic emission, not purely by the change
of the absorber as discussed by \citet{Mol07} in the frame work of a
``patch torus'' model \citep{Eli06}. While there remains a possibility
that Compton thick absorber ($N_{\rm H} > 10^{24}$ cm$^{-2}$) may
completely cover a part of the emission region to cause the flux
variability, it would produce signals of heavy obscuration such as a
deep iron-K edge feature in the X-ray spectra, which are not seen in
the data. Thus, at least variability of the continuum flux is required
to explain these results, while that of the absorber can also
contribute to the variability, in particular below 10 keV
(\citealt[see e.g.,][]{Ris05} for NGC~1365).

We significantly detect an iron-K emission line and obtain a tight
constraint on the reflection component, even though it is quite weak
as reported by \citet{Mol07}. The reflection strength, $R \simeq 0.2$
is much smaller compared with Seyfert 1 galaxies, which typically have
$R \sim 1$ \citep{Dad08}. Even correcting for a variability effect
that the direct continuum flux was higher in our observations by a
factor of $\approx$1.4 than the 22-months average, the
reflection is still small, $R\simeq 0.3$. The weak reflection is
also consistent with the narrow iron-K emission line, which indicates
that the reflection is mainly produced by relatively outer parts of
the disk (hence with a small solid angle), unlike the results from
typical Seyfert 1 galaxies \citep{Dad08}. Our 4C~50.55 result confirms
the trend reported by \citet{Sam02} for radio loud AGNs.

The analysis of the iron-K line profile yields an (apparently) large
innermost radius, $r_{\rm in} \sim 700 r_{\rm g}$, by assuming
an emissivity law of $r^{-3}$. Our diskline result suggests it
unlikely that a ``standard disk'' extends down to close to the
innermost stable circular orbit (ISCO) around the black hole, $<$ 6
$r_{\rm g}$. Instead, it is possible that the disk is there but its
inner part is covered by an optically thick corona, as estimated by
our Comptonization model fit, which would smear out relativistic
iron-K broad lines. Note that the obtained $r_{\rm in}$ value does not
directly mean that the disk is truncated at that radius, because (1)
the estimated $r_{\rm in}$ critically depends on the emissivity
profile and (2) there may be another line component from distant
parts, such as the torus. If the scale height of the X-ray irradiating
corona is sufficiently small, then one would expect a flatter slope
for the emissivity law, even close to $r^{-2}$. In this extreme case,
we were not able to obtain good constraints on $r_{\rm in}$ from our
data. To examine the second possibility, we apply a two-components
line model to the XIS 3--9 keV spectra, consisting of a narrow
Gaussian at 6.4 keV and a broad diskline, which represents that from
the torus and disk, respectively. We obtain a worse fit than the
single diskline fit by $\Delta \chi^2 = 5$ even with a larger degrees
of freedom, suggesting that the two components model is not a good
description of the data. Nevertheless, when $r_{\rm in}$ of the disk
line component is fixed at 10 $r_{\rm g}$ as found from 3C~120
\citep{Kat07}, both lines are found to be significant with equivalent
widths of 19$\pm 6$ eV and 24$\pm$15 eV,
respectively. Thus, we do not completely exclude the possibility for
the presence of a moderately broadened iron-K line in the observed
spectra of 4C~50.55.

The inferred geometry of the accretion disk in 4C~50.55 (i.e.,
truncated and/or inner parts covered by corona) may be common features
of AGNs with powerful jets. Recent {\it Suzaku} studies indicate that
radio galaxies also have relatively narrow iron-K emission lines e.g.,
$r_{\rm in} >$ 20 $r_{\rm g}$ for 3C 390.3 \citep{Sam09} and $r_{\rm
in} >$ 44 $r_{\rm g}$ for 4C +74.26 \citep{Lar08} from the single
diskline fit, and $r_{\rm in} = (9\pm1) r_{\rm g}$ for 3C 120 from the
multiple components fit \citep{Kat07}. This result is in accordance
with an expectation from theories that jets are more easily produced
by radiatively inefficient accretion flow than by a standard disk.

Another key parameter to understand the accretion flow is the
Eddington ratio, which is estimated to be $L_{\rm bol}/L_{\rm Edd}
\sim 0.4$ for 4C 50.55 (section~\ref{differ_SED}). Similarly, we
also estimate that of 3C~120 to be $L_{\rm bol}/L_{\rm Edd} \sim 0.5$,
using the 2--10 keV flux \citep{Kat07} and the black hole mass of
$10^{7.7}$ \solarmass\ \citep{Pet04}. Thus, these two sources may
belong to a very similar class of AGNs, except for the radio loudness
to the X-ray flux (log $R_{\rm X} = -2.1$ for 3C~120 and log $R_{\rm
X} = -3.6$ for 4C~50.55), which could be partially explained by
the small inclination angle of 3C~120 ($i<14^{\circ}$; see
\citealt{Kat07}) compared with 4C~50.55 ($i \sim 35^{\circ}$). 
The physical reason for the difference in their X-ray spectra that the
reflection component and iron-K lines are stronger in 3C~120
($R\sim0.7$) is not clear at present.

4C 50.55 and 3C~120 are rare objects having distinctively high
fractions of Eddington luminosity compared with other typical BLRGs,
for instance, $L_{\rm bol}/L_{\rm Edd}$ = 0.01--0.07 for 3C 390.3
\citep{Sam09, Lew06}, $\sim 0.04$ for 4C~+74.26 \citep{Lar08}, and
0.001--0.002 for Arp 102B \citep{Lew06}. By analogy to the Galactic
black holes, these low Eddington ratio sources likely correspond to
the low/hard state, where the accretion disk is accompanied by steady
jets, while normal Seyfert galaxies may do to the high/soft state,
where the disk extends close to the ISCO with quenched jet
activity. The accretion flows in 4C~50.55 and 3C~120 could be
explained as a high luminosity end of the low/hard state.
Alternatively, they may be another state achieved with even higher
mass accretion rates than in the high/soft state, where the disk
structure is also similar to that found in the low/hard state (i.e.,
truncated disk). For this possibility, it is interesting to note the
similarly to the high-Eddington ratio Galactic black hole
GRS~1915+105, which exhibits a similarly narrow iron-K emission line
over a Comptonization dominated continuum, implying that the inner
disk is fully covered by a corona \citep{Ued10}; in GRS~1915+105, a
compact jet is also detected in a steady state with a hard spectrum,
so-called in Class~$\chi$ \citep[see e.g.,][]{Fen04a}. In summary, the
unified picture of accretion flows over a wide range of black hole
mass is far from established. Further systematic studies of the
accretion disk structure of radio loud AGNs at various accretion rates
based on detailed X-ray spectroscopy and multi-wavelengths data are
very important to reveal these fundamental problems.

\acknowledgments

We thank Gerry Skinner for providing the {\it Swift}/BAT light curve
of 4C~50.55, and the {\it Suzaku} team for the calibration of the
instruments. Part of this work was financially supported by
Grants-in-Aid for Scientific Research 20540230 (YU) and 20740109 (YT),
and by the grant-in-aid for the Global COE Program ``The Next
Generation of Physics, Spun from Universality and Emergence'' from the
Ministry of Education, Culture, Sports, Science and Technology (MEXT)
of Japan.


\begin{thebibliography}{}
\bibitem[Abdo et al.(2009)]{Abd09} Abdo, A. A., et al.\ 2009, \apjs, 183, 46
\bibitem[Anders \& Grevesse(1989)]{and89} Anders, E. \& Grevesse, N. 1989, Geochimica et Cosmochimica Acta, 53, 197
\bibitem[Bird, Barlow \& Bassani(2004)]{Bir04} Bird, A. J. et al.\ 2004, \apj, 607, 33 
\bibitem[Dadina(2008)]{Dad08} Dadina, M.\ 2008, \aap, 485, 417
\bibitem[Elitzur \& Shlosman(2006)]{Eli06} Elitzur M., \& Shlosman, I.\ 2006, \apj, 648, 101
\bibitem[Elvis et al.(1994)]{Elv94} Elvis, M., Wilkes, B. J., McDowell, J. C., Green, R. F., Bechtold, J., 
Willner, S. P., Oey, M. S., Polomski, E., \& Cutri, R.\ 1994, \apjs, 95, 1
\bibitem[Fabian et al.(1989)]{Fab89} Fabian, A. C., Rees, M. J., Stella, L., 
\& White, N. E.\ 1989, \mnras, 238, 729
\bibitem[Fender \& Belloni(2004)]{Fen04a} Fender, R. \& Belloni, T. 2004, \araa, 42, 317
\bibitem[Fender et al.(2004b)]{Fen04b} Fender, R. P., Belloni, T. M., \& Gallo, E.\ 2004, \mnras, 355, 1105
\bibitem[Fukazawa et al.(2008)]{Fuk08} Fukazawa, Y., Mizuno, T., Takahashi, H.,
Enoto, T., Kokubun, M., Watanabe, S., \& the HXD team\ 2008, JX-ISAS-SUZAKU-MEMO-2008-01 
\bibitem[Giovannini et al.(1988)]{Gio88} Giovannini, G., Feretti, L., Gregorini, L., \& Parma, P.\ 1988, \aap, 199, 73
\bibitem[Grandi et al.(2006)]{Gra06} Grandi, P., Malaguti, G., \& Fiocchi, M.\ 2006, \apj, 642, 113
\bibitem[Gruber et al.(1999)]{Gru99} Gruber, D. E., Matteson, J. L., Peterson, L. E., \& Jung, G. V.\ 1999, \apj, 520, 124
\bibitem[Grupe et al.(2010)]{Gru10} Grupe, D., Komossa, S., Leighly, K. M., \& Page K. L.\ 2010, \apjs, 187, 64
\bibitem[Ishisaki et al.(2007)]{Ish07} Ishisaki, Y., et al.\ 2007, \pasj, 59, 113
\bibitem[Kalberla et al.(2005)]{Kal05} Kalberla, P. M. W., Burton, W. B., Hartmann, D., Arnal, E. M., Bajaja, E., Morras, R., P\"{o}ppel, W. G. L.\ 2005, \aap, 440, 775
\bibitem[Kataoka et al.(2007)]{Kat07} Kataoka, J., et al.\ 2007, \pasj, 59, 279 
\bibitem[Komatsu et al.(2009)]{Kom09} Komatsu, E., et al.\ 2009, \apj, 180, 330
\bibitem[Kubo et al.(1998)]{Kub98} Kubo, H., Takahashi, T., Madejski, G., Tashiro, M., Makino, F., Inoue, S., 
\& Takahara, F.\ 1998, \apj, 504, 693
\bibitem[Larsson et al.(2008)]{Lar08} Larsson, J., Fabian, A. C., Ballantyne, D. R., \& Miniutti, G.\ 2008, \mnras, 388, 1037
\bibitem[Lewis \& Eracleous(2006)]{Lew06} Lewis, K. T., \& Eracleous, M.\ 2006, \apj, 642, 711 
\bibitem[Lubinski et al.(2010)]{Lub10} Lubinski, P., Zdziarski, A. A., Walter, R., Paltani, S., Beckmann, V., Soldi, S., Ferrigno, C., \& Courvoisier, T. J.-L.\ 2010, arXiv:1005.0842
\bibitem[Maeda et al.(2008)]{Mae08} Maeda, Y., et al.\ 2008, JX-ISAS-SUZAKU-MEMO-2008-06
\bibitem[Magdziarz \& Zdziarski(1995)]{Mag95} 
Magdziarz, P. \& Zdziarski, A. A.\ 1995, \mnras, 273, 837
\bibitem[Massetti et al.(2004)]{Mas04} Massetti, N., Palazzi, E., 
Bassani, L., Malizia, A., \& Stephen, J. B.\ 2004, \aap, 426, 41
\bibitem[Mitsuda et al.(2007)]{Mit07} Mitsuda, K., et al.\ 2007, \pasj, 59, S1
\bibitem[Molina et al.(2007)]{Mol07} Molina, M., et al.\ 2007, \mnras, 382, 937
\bibitem[Pandey et al.(2006)]{Pan06}
Pandey, M., Manchanda, R. K., Rao, A. P., Durouchoux, P., \& Ishwara-Chandra\ 2006, \aap, 446, 471
\bibitem[Peterson et al.(2004)]{Pet04} Peterson, B. M., et al.\ 2004, \apj, 613, 682
\bibitem[Risaliti et al.(2005)]{Ris05} Risaliti, G., Elvis, M., Fabbiano, G., Baldi, A., \& Zezas, A.\ 2005, \apj, 623, 93
\bibitem[Sambruna et al.(2009)]{Sam09} Sambruna, R. M., et al.\ 2009, \apj, 700, 1473
\bibitem[Sambruna et al.(2002)]{Sam02} Sambruna, R. M., Eracleous, M., \& Mushotzky, R. F.\ 2002, New A Rev., 46, 215
\bibitem[Sunyaev \& Titarchuk(1980)]{Sun80} Sunyaev, R. A., \& Titauchuk, L. G.\ 1980, \aap, 86, 121
\bibitem[Tueller et al.(2008)]{Tue08} Tueller, J., et al.\ 2008, \apj, 681, 113
\bibitem[Terashima \& Wilson(2003)]{Ter03} Terashima, Y., \& Wilson, A. S.\ 2003, \apj, 583, 145
\bibitem[Ueda et al.(2010)]{Ued10} Ueda, Y., et al.\ 2010, \apj, 713, 257
\bibitem[Vestergaard \& Peterson(2006)]{Ves06} Vestergaard, M., \& Peterson, B. M.\ 2006, \apj, 641, 689
\bibitem[Winter et al.(2010)]{Win10} Winter, L. M., Lewis, K. T., Koss, M., Veilleux, S., Keeney, B., 
\& Mushotzky, R. F.\ 2010, \apj, 710, 503
\bibitem[Zycki et al.(1999)]{Zyc99} Zycki, P. T., Done, C., \& Smith, D. A.\ 1999, \mnras, 309, 561
\end{thebibliography}
\end{document}